\documentclass{emulateapj} 

\usepackage{ifthen}
\newboolean{color}

\setboolean{color}{false}
\setboolean{color}{true}

\slugcomment{\\Submitted to ApJ}

\usepackage{amsmath}
\usepackage{graphicx}
\usepackage{float}
\usepackage{amssymb,amsmath}
\usepackage{epsfig,floatflt}
\usepackage{subfigure}

\begin{document}

\newcommand \thi{\thinspace}
\newcommand \noi{\noindent}

\long\def\jumpover#1{{}}

\def\ni{{\noindent}}
\def \at{{\rm\char'100}}
\newcommand \eg{{{\it e.g.},\ }}
\newcommand \etal{{\it et al.\ }}
\newcommand \etc{{\it etc.\ }}
\newcommand \cf{{\it cf.\ }}
\newcommand \ia{{{\it inter alia},\ }}
\newcommand \ie{{{\it i.e.},\ }}
\newcommand \via{{\it via\ }}
\newcommand \viz{{\it viz.\ }}
\newcommand \vs{{\it vs.\ }}

\newcommand \Teff{{$T_{\rm {ef\!f}} $}}
\newcommand \teff{{T_{\rm {ef\!f}} }}
\newcommand \Lo{{$L_\odot $}}
\newcommand \Mo{{$M_\odot $}}
\newcommand \Log{{\rm Log}\thi}
\newcommand \approxgt{\,\raise2pt \hbox{$>$}\kern-8pt\lower2.pt\hbox{$\sim$}\,}
\newcommand \approxlt{\,\raise2pt \hbox{$<$}\kern-8pt\lower2.pt\hbox{$\sim$}\,}
\newcommand\dotd{\hbox{$.\!\!^{\rm d}$}}
\newcommand\dotm{\hbox{$.\!\!^{\rm m}$}}
\newcommand\dd{{$^d$}}
\newcommand\ML{{$M$-$L$\ }}
\newcommand\aanda{A\&A}
\newcommand\RV{$V_r$}
\newcommand\DP{{$\Delta(\Delta\Phi_1)$}}
\newcommand\DDP{{$|\Delta(\Delta\Phi_1)|$}}
\newcommand\dinv {{\thinspace d$^{-1}$}}

\newcommand\MB{{M$_{\rm B}$}}
\newcommand\MR{{M$_{\rm R}$}}
\newcommand\OI{{OGLE I}}
\newcommand\OV{{OGLE V}}

\title{Ultra-low Amplitude Variables in the LMC -- Classical Cepheids, Pop. II
Cepheids, RV~Tau Stars and Binary Variables }
\author{J. Robert Buchler\altaffilmark{1,2},
Peter R. Wood\altaffilmark{2}
\& Igor Soszy{\'n}ski\altaffilmark{3}}
\altaffiltext{1}{\ni Physics Department, University of Florida,
Gainesville, FL 32611, USA; buchler\at phys.ufl.edu}
\altaffiltext{2}{\ni Research School
of Astronomy \& Astrophysics, Australian National University, Canberra,
AUSTRALIA, wood\at mso.anu.edu.au}
\altaffiltext{3}{Warsaw University Observatory, Al. Ujazdowskie 4, 00-478
Warszawa, POLAND, soszynsk\at astrouw.edu.pl}

\begin{abstract}

A search for variable stars with ultra-low amplitudes (ULA), in the millimag
range, has been made in the combined MACHO and OGLE data bases in the broad
vicinity of the Cepheid instability strip in the HR diagram.  A total of 25
singly periodic and 4 multiply periodic ULA objects has been uncovered.  Our
analysis does not allow us to distinguish between pulsational and ellipsoidal
(binary) variability, nor between LMC and foreground objects.  
However, the objects are strongly clustered and appear
to be associated with the pulsational instability strips of LMC Pop. I and II
variables.  \\
When combined with the ULA variables of \cite{bwks05} a total of
 20 objects fall close to the
classical Cepheid instability strip.  However, they appear to fall on parallel
period-magnitude relations that are shifted to slightly higher magnitude which
would confer them a different evolutionary status.  \\ 
Low amplitude RV~Tauri and Pop. II Cepheids have been uncovered that do not
appear in the MACHO or OGLE catalogs.  Interestingly, a set of binaries seem to
lie on a PM relation that is essentially parallel to that of the
RV~Tauri/Pop.~II Cepheids.
 
\end{abstract}


\section{Introduction}

In a previous paper \citep{bwks05}, we searched for periodic variability in the
light curves of selected LMC stars in MACHO Field 77.  The stars were required
to lie in and around the Cepheid instability strip in the HR diagram.  We found
seven stars with a periodic behavior of less than 0.006 mag. and labeled them
ultra-low amplitude (ULA) Cepheids.

In this paper, we extend our analysis to all stars that are {\sl common} to the
OGLE II and III \citep{ogle} and the MACHO \citep{macho99} data bases.  The
stellar selection is defined in terms of the OGLE stars that are located in a
parallelogram in the HR diagram defined by $14 < V < 17.5$ and $18.8 < V + 9.18
\thi (V-I) <25.0$.  This region has been chosen by visual inspection to include
the instability strip and colors 0.15 blueward and redward.  It contains a
mixture of non-oscillatory giants of spectral type F, and variable stars such
as Cepheids, W~Vir stars, and ellipsoidal variables and other binaries.  The
region has been converted from Johnson $V$ and Cousins $R_{\rm c}$ into MACHO
blue magnitude (\MB) and red magnitude (\MR) using the transforms given in
Alcock \etal (1999).  Position-matched stars were examined in both the 
MACHO and OGLE data bases.

The standard LMC Cepheids have already been identified both in the MACHO, EROS
and the OGLE data bases separately for amplitudes greater than about 0.01
mag. \citep{beaulieu,welch,ogle, kanbur}.  Our goal here is to concentrate on
ULA objects.

\begin{figure*}
\epsscale{1.1}
\epsscale{0.95}
\epsscale{0.85}
\begin{center}
 \plotone{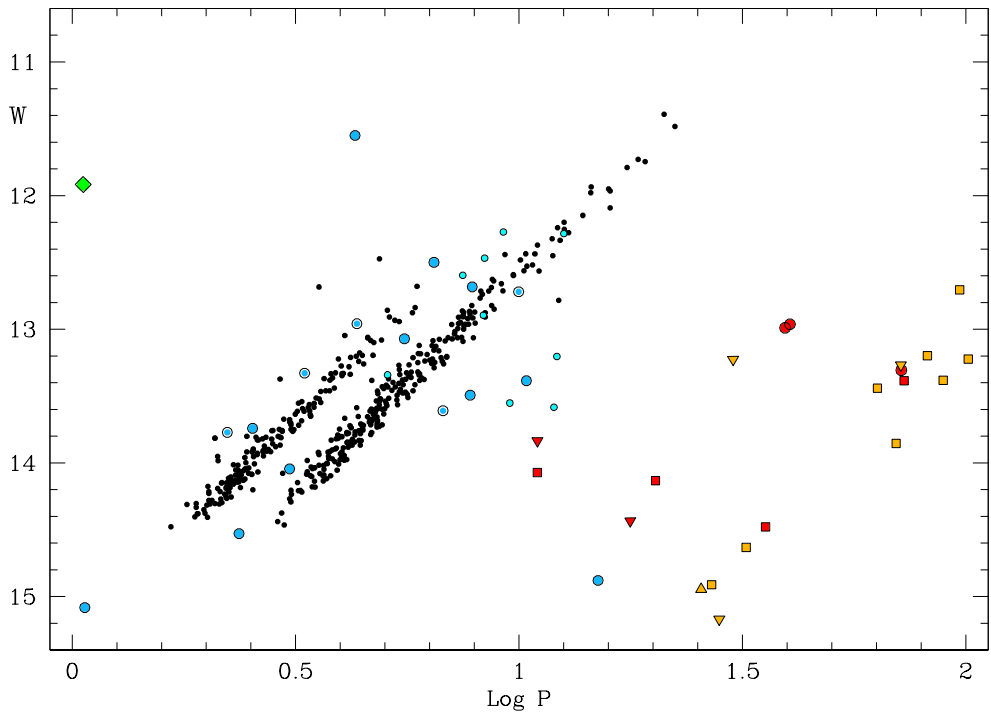}
\caption{
Period--$W$ plot for all singly periodic variables of all amplitudes:
  ULA variables (light blue and cyan),
 Classical Cepheids (black dots);
 other larger amplitude objects categorized by light curve shape: 
 non alternating (upside down triangles),
 regularly alternating (squares),  semiregular alternations (circles), 
 ULA alternating (triangle); Pop. II variable stars (red),  binaries
 (orange);  
 special object (green diamond).  See text for further details.\\
}
\label{1}
\end{center}
\end{figure*}

\begin{table*}
\caption{\small Singly Periodic ULA Objects}
Columns:  OGLE and MACHO IDs, I band mag, color, period, frequency, \MR\ and \OI\ band
Fourier amplitudes.  See text for cols. 9 and 10. 
Object SC2 357794
lies considerably above the Cepheid P--M relation ($W$,\,$\Log P$) and
is a strange Cepheid candidate.
\vspace{-0mm}
\begin{center}
\medskip
\begin{tabular}{l l c l r c c c r r}
\hline\hline
\noalign{\smallskip}
 OGLE & MACHO & I & (V-I) & $P_R$ & $f_R$ & $A_R$ &$A_I$ & &  \\
\noalign{\smallskip} 
\hline   
\hline
\noalign{\smallskip}
SC6  57397  &   78.6463.64  &15.220 &0.444 & 2.3629 &0.42320   &0.0102&  0.0091 & RBI \\ 
SC8  76141  &   79.5746.15  &13.994 &0.846 & 7.8600 &0.12723   &0.0090&  0.0057 & RBI \\
SC11 325504 &   79.4779.17  &13.772 &0.821 & 6.4537 &0.15495   &0.0027&  0.0028 & RBI \\
SC15 5854   &   17.3199.23  &15.496 &0.267 & 1.0674 &0.93685   &0.0080&  0.0115 & RBI \\
SC13 223889 &    1.4295.12  &14.452 &0.891 & 5.5415 &0.18046   &0.0030&  0.0039 & RBI \\
SC1  335392 &   81.8881.40  &14.679 &0.835 &10.3895 &0.09625   &0.0036&  0.0028 & RBI \\
SC1  330647 &   81.8758.16  &15.307 &0.276 & 15.029 &0.06654   &0.0036&  0.0030 & RBI \\
SC13 178852 &  19.4299.298  &14.896 &0.745 & 2.5337 &0.39467   &0.0027 & 0.0038 & RI &x \\
SC5  416555 &   77.7189.9   &14.161 &0.431 & 7.7750 &0.12862   &0.0026 & 0.0019 & RI &x \\
SC15 196983 &  1.3568.30    &14.855 &0.523 & 3.0663 &0.32613   &0.0029 & 0.0016 & BI &x \\
\noalign{\smallskip}
SC2  6920   &    7.8026.17  &13.946 &0.871 & 7.4876 &0.13355   &0.0042&  0.0035 & RBI & $\dagger$\\
SC19 64022  &  12.10317.10  &13.783 &0.975 & 9.2242 &0.10841   &0.0033&  0.0043 & RBI & $\dagger$\\
SC9  127569 &   79.5378.48  &14.818 &0.817 & 9.5378 &0.10485   &0.0035&  0.0029 & RBI & $\dagger$\\
SC11 300879 &    79.4775.9  &13.654 &0.884 & 12.600 &0.07937   &0.0063&  0.0067 & RBI & $\dagger$\\
SC14 7197   &     1.3561.6  &14.580 &0.799 & 5.0807 &0.19682   &0.0048&  0.0045 & RBI & $\dagger$\\
SC7  344553 &   80.6351.14  &13.886 &0.915 & 8.3764 &0.11938   &0.0049&  0.0025 & RBI & $\dagger$\\
SC20 181340 &  12.10922.11  &14.873 &0.832 & 11.970 &0.08354   &0.0049&  0.0040 & RBI & $\dagger$\\
SC21 140323 &    6.6697.11  &14.136 &0.801 & 8.3250 &0.12012   &0.0041&  0.0030 & RBI & $\dagger$\\
SC13 191166 & 19.4302.323   &14.450 &0.804 &12.1561 &0.08226   &0.0032&  0.0029 & RBI & x\thi $\dagger$\\
\noalign{\smallskip}
SC2  357794 &    81.8392.8  &13.109 &1.006 & 4.2964 &0.23275   &0.0059 & 0.0013 &S RBI\\
\noalign{\smallskip}
SC4 296029 & 77.7428.36  & 14.819 &  0.676 & 2.2250 & 0.44953  &0.0024 & 0.0022 &RBI&*\\
SC4 323401 & 77.7430.18  & 14.505 &  0.759 & 3.3116 & 0.30187  &0.0042 & 0.0033 &RI &*\\
SC3 153959 & 77.7789.25  & 14.116 &  0.901 & 9.9817 & 0.10018  &0.0045 & 0.0034 &RBI&*\\
SC3 35239  & 77.7668.981 & 14.149 &  0.768 & 4.3373 & 0.23056  &0.0054 & 0.0032 &RI &*\\
SC4 176301 & 77.7306.43  & 14.566 &  0.617 & 6.7609 & 0.14791  &0.0109 & 0.0025 &RBI&*\\
\noalign{\smallskip}
\noalign{\smallskip}
\hline
\hline
\end{tabular}
\label{tab:ula}
\end{center}
\end{table*}

\begin{figure}
\epsscale{1.15}
\begin{center}
 \plotone{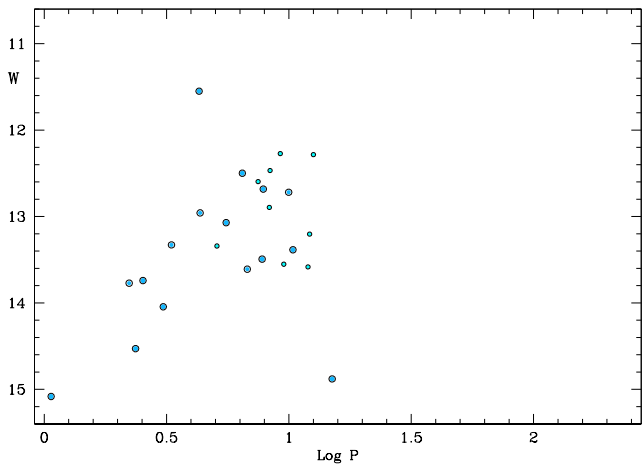}
\caption{Top: 
Period--$W$ plot  of all singly periodic ULA variables in 
the searched period range, showing a clear clustering in the Cepheid
instability strip. 
}
\label{2}
\end{center}
\end{figure}

Fourier analysis is known to be very good at detecting periodicity in data sets
even in the presence of large noise.  We have therefore performed a Fourier
analysis of the light curves in the combined data base in the three bands MACHO
\MR\ and \MB\ and OGLE I with the MUFRAN code (multi-frequency analysis,
Koll\'ath 1990).  The \OV\ band data set is generally sparse and less useful in
the detection of ULA variables.  The analysis is performed in the 0 -- 0.99\thi
d$^{-1}$ frequency range, but the plotted frequency range in the subsequent
figures is 0.004 -- 0.99 d$^{-1}$, so as to avoid both the yearly and the daily
aliases.  One has to be careful not to miss a potential longer period that can
occur in some ellipsoidals \citep{menni}.  We have sifted obvious outlying
points from the light curves prior to analysis.  In this paper all quoted
amplitudes refer to Fourier amplitudes.  They are determined with MUFRAN with a
sinusoidal linear least-squares fit to the light curve using the frequency from
the dominant Fourier peak.

We first reduce the set of over 3000 objects to those in which there are
coincidences among the 8 largest Fourier peaks in \MR\ and \OI.  For the
standard, large amplitude Cepheids the peaks are very sharp, and these variable
stars are thus readily identified.  In the case of the ULA stars, the Fourier
peaks \MR\ amplitudes fall well below 0.01\thi mag, but nevertheless there
are coincidences among the \MR\ and I peaks. Each of these cases has to be
examined individually to ascertain that the detected variability is not
spurious.  For this purpose we also use the \MB\ data.

We mention in passing that the typical \MR, \MB\ and \OI\ window functions have
a very large sharp central peak with yearly side peaks at least 45\% lower, but
otherwise are very dull.
 
We note one peculiarity with the Fourier spectra which is a result of the
extremely large 1 day alias of the data: all significant peaks of frequency $f$
occur with an alias at $1 - f$, \ie with peaks placed symmetrically around the
frequency 0.5\thi d$^{-1}$.  The peaks can be of comparable amplitude.  The
ambiguity of the period determination can generally be resolved by a comparison
of the peaks in \MR, \MB, \OI\ and V.  When the symmetrically placed peak is
absent, which happens when the original peak is barely above the noise, it
likely is an indication that the frequency is spurious.

The objects we have found have been grouped into a singly periodic and a
multiperiodic set, and will present them separately.


\section{Singly Periodic Objects}

Figure~\ref{1} displays {\sl all the singly periodic} variables
that we have uncovered in a period--magnitude (PM)
diagram.  $W= I - 1.55 \thi (V-I)$ is the Wesenheit function, a reddening-free 
magnitude estimate.  We
now attempt to identify the nature of these objects.

The black dots in Fig.~\ref{1} fall on two parallel bands which
correspond to the fundamental and overtone classical Cepheids.  They are
included in the figures for reference purposes.  All these Cepheids have I band
amplitudes greater than $\sim$0.015\thi mag.

The red and orange symbols below the classical Cepheids have relatively large
amplitudes.  We will return to them below.

\begin{figure*}
\begin{center}
\epsscale{0.67}
{\plotone{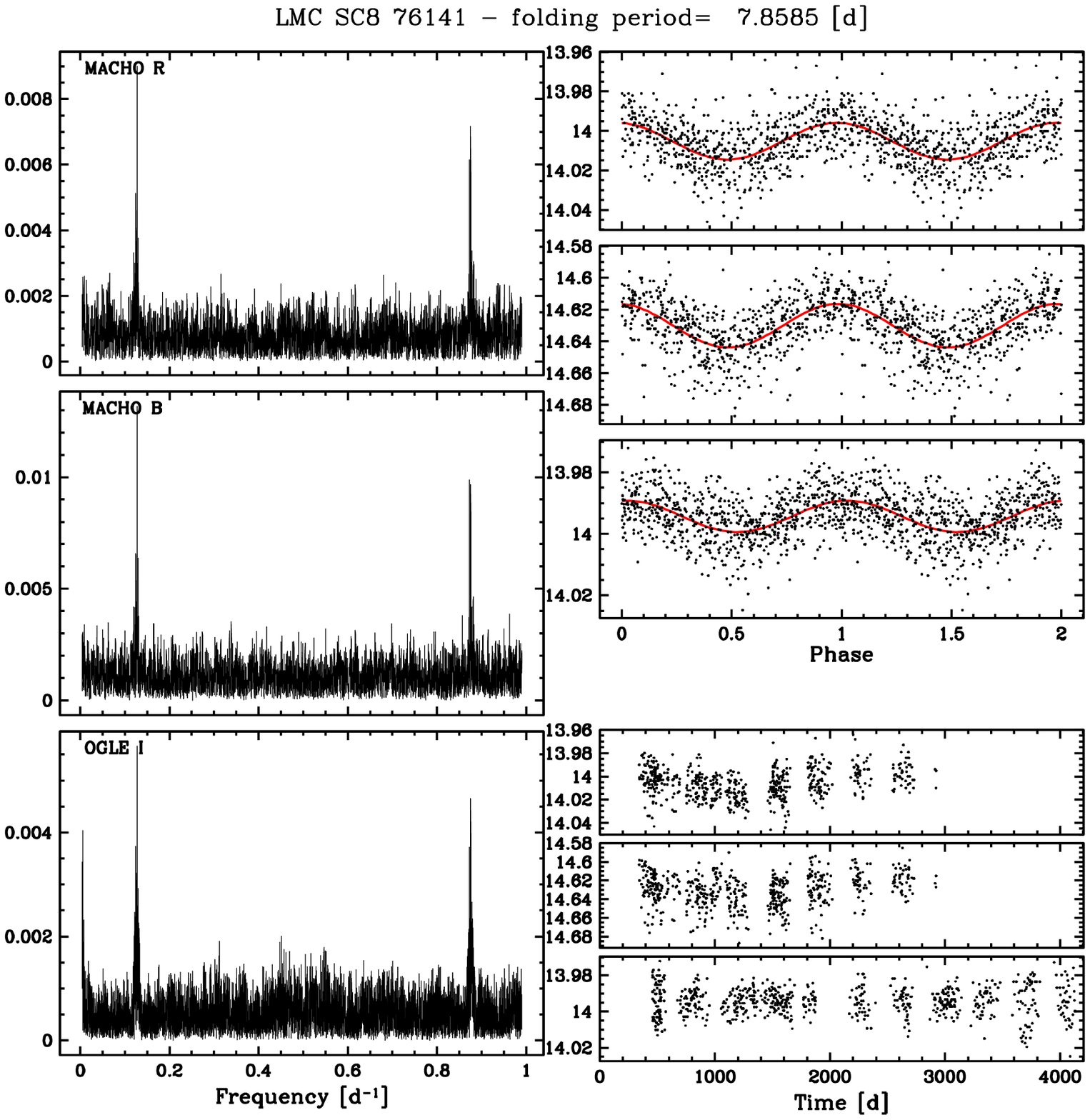}}
\caption{LMC SC8 76141 --
Left side, top down: 
Fourier amplitude spectra of the \MR, \MB\ and \OI\ data;
Right side, top: folded and phased data;
folding period appears on top;
bottom: corresponding data points.
}
\label{3}
\end{center}
\end{figure*}

\begin{figure*}
\begin{center}
\epsscale{0.67}
{\plotone{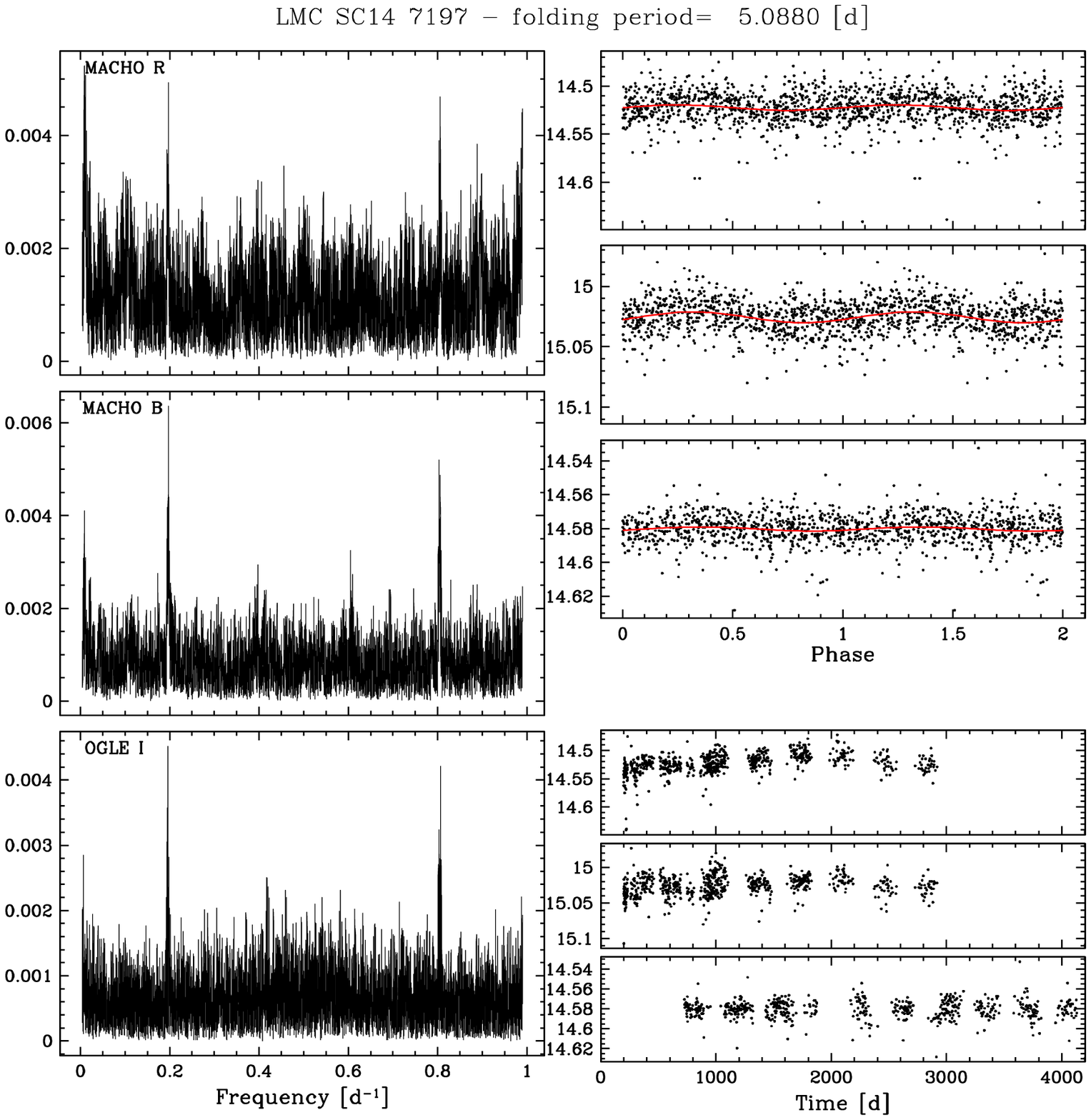}}
\caption{LMC SC14 7197 --
Left side, top down:
Fourier amplitude spectra of the \MR, \MB\ and \OI\ data;
Right side, top: folded and phased data;
folding period appears on top;
bottom: corresponding data points.
}
\label{4}
\end{center}
\end{figure*}

\begin{figure*}
\begin{center}
\epsscale{0.67}
{\plotone{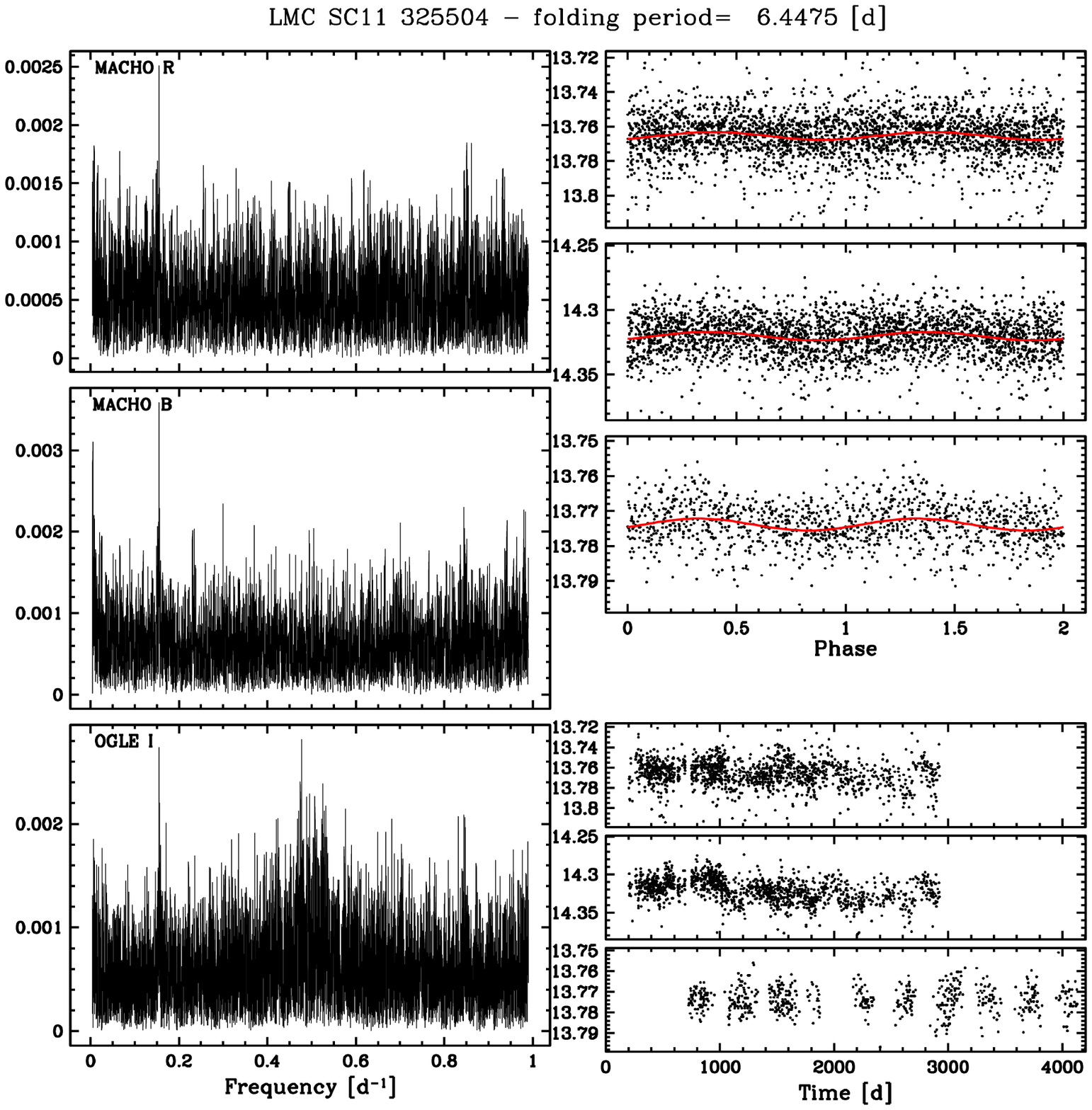}}
\caption{LMC SC11 325504 --
Left side, top down: 
Fourier amplitude spectra of the \MR, \MB\ and \OI\ data;
Right side, top: folded and phased data;
folding period appears on top;
bottom: corresponding data points.
}
\label{5}
\end{center}
\end{figure*}

\begin{figure*}
\begin{center}
\epsscale{0.67}
{\plotone{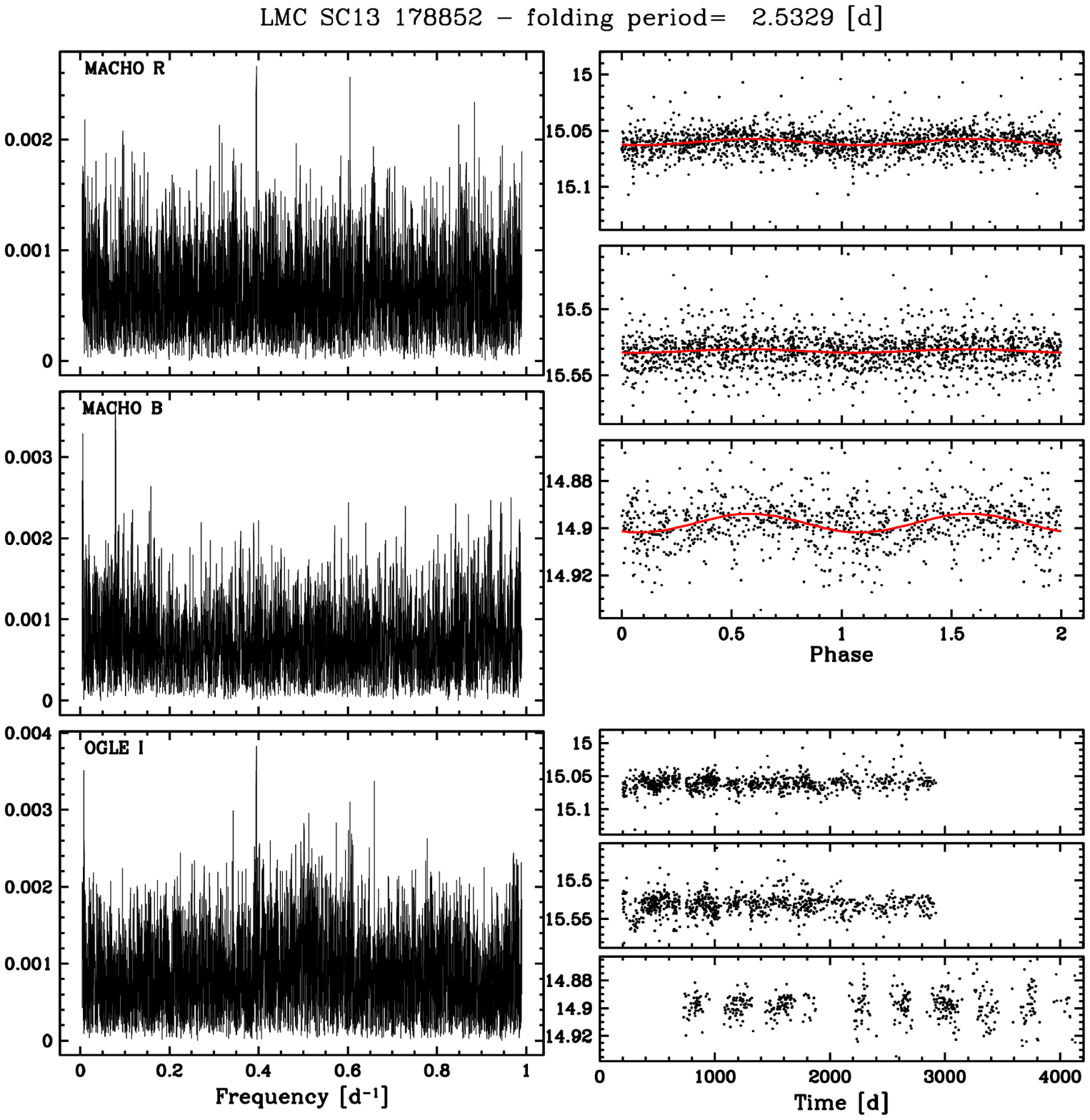}}
\caption{LMC SC13 178852 --
Left side, top down:
Fourier amplitude spectra of the \MR, \MB\ and \OI\ data;
Right side, top: folded and phased data;
folding period appears on top;
bottom: corresponding data points.
}
\label{6}
\end{center}
\end{figure*}

\begin{figure*}
\begin{center}
\epsscale{0.67}
{\plotone{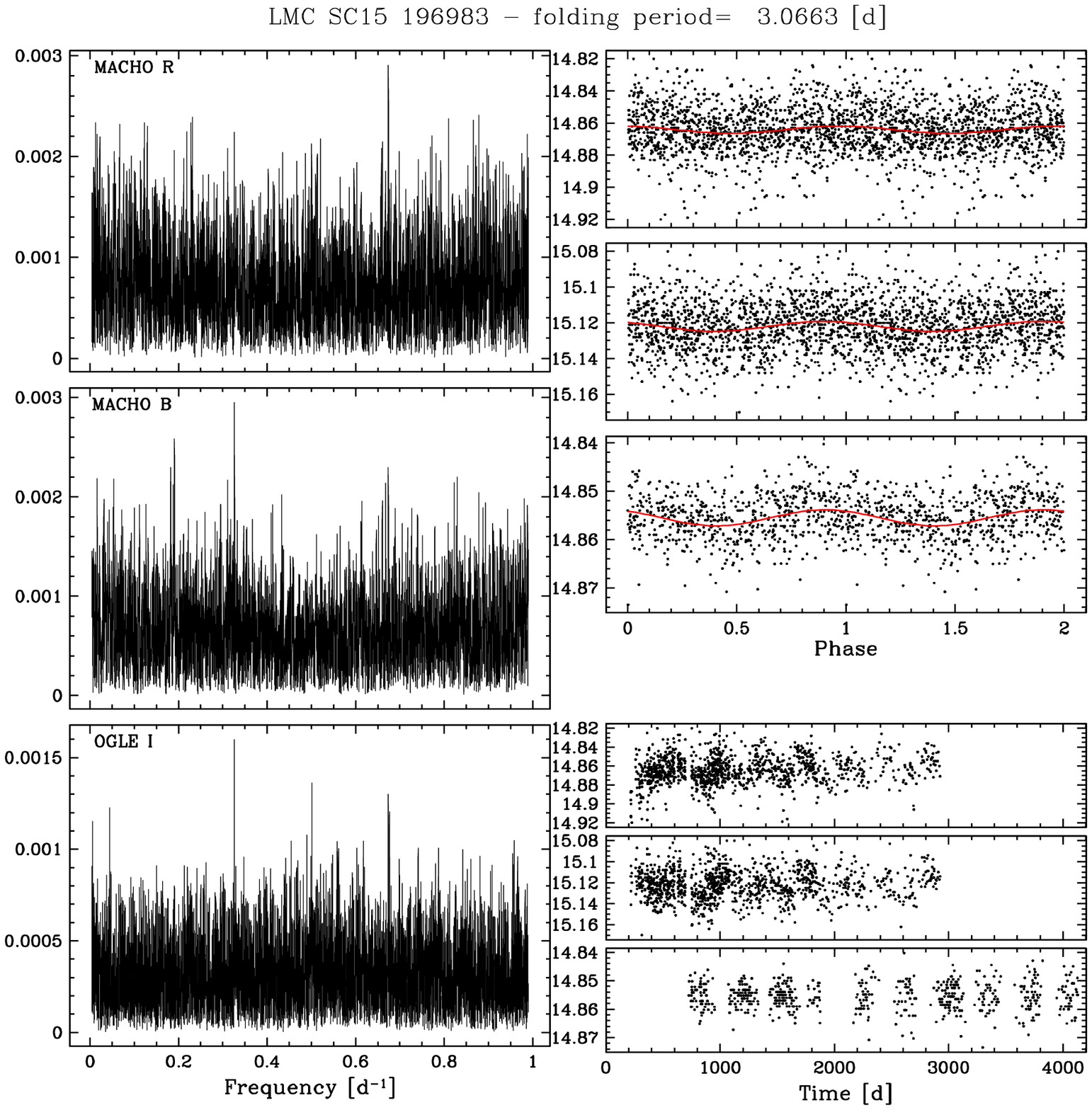}}
\caption{LMC SC15~196983 --
Left side, top down: 
Fourier amplitude spectra of the \MR, \MB\ and \OI\ data;
Right side, top: folded and phased data;
folding period appears on top;
bottom: corresponding data points.
}
\label{7}
\end{center}
\end{figure*}

The 25 ULA variables we have found are listed in Table~\ref{tab:ula}.  All
objects have I band amplitudes less than 0.0067, except for SC15~5854 which has
an amplitude of 0.0115.  The first two columns of Table~\ref{tab:ula} give the
MACHO and OGLE names, respectively, followed by the average I magnitude and
(V--I) color, and the period, $P_R$, frequency $F_R$, and the Fourier
amplitudes in \MR\ and in \OI.  The frequency $f_R$ (and period) is optimized
in a linear squares Fourier fit to the \MR\ data starting with the highest \MR\
spectral peak.

The penultimate column denotes the bands in which there is a strong common peak
in the Fourier spectra.  By strong we mean that its amplitude stands out by at
least a factor of two above the noise level.  The asterisk in the last column
denotes stars that were already discussed by \cite{bwks05}.

We have plotted the ULA variables as light blue and cyan circles.
For the four ULA objects that are branded
with an X, the evidence for very low amplitude periodicity is a little less
strong in the sense that it does not satisfy all our criteria.  We will return
to them shortly.

Because of the observational noise we have no way of distinguishing ULA
pulsating stars from ellipsoidal variables.  However, we shall argue that there
are very few ellipsoidals in our sample.  For that purpose we present
Figure~\ref{2} which locates, unencumbered, {\sl only} the
ULA variables in the whole range we have analyzed, namely the $W$ range
corresponding to the chosen parallelogram in the HR diagram (see \S1) and the
frequency range from 0.004 to 0.99\thi d$^{-1}$, \ie $\Log P$[d] = 0.0044 --
2.40.  The way the ULA objects cluster near the instability strip in the PM
diagram suggests that a large majority of the objects are
intrinsic variables and are associated with either the classical Cepheids or,
the lower lying ones, the Pop. II Cepheids.

In the OGLE data \citep{ogle08} each frame was processed with the DoPhot
photometry program. In this way independent PSF profile photometry of each
object was derived, supplemented with astrometric information (current X,Y
position) obtained via PSF fitting.  As a result proper motions can be
inferred.  In fact, unpublished OGLE-III data suggests that 9 objects show
proper motion, which in turn suggests that they are foreground stars.  They are
represented as smaller cyan filled circles in Figs.~\ref{1} and \ref{2}, and
are branded with a dagger in column 10 of Table~\ref{tab:ula}.  However, if
they are indeed foreground stars in the solar vicinity that overlap in the CMD
with the LMC instability strip \citep{sos02} it is nevertheless curious that 6
of them, additionally, appear so near the LMC classical Cepheid PM relations,
and the other 3 cluster just below.

If they are foreground stars, one would think that they are some sort of
main-sequence ellipsoidal binary in the solar vicinity.  Generally, spectra can
be used to tell giants from dwarfs and to derive radial velocities, so potentially 
foreground or LMC status could be investigated by spectral observations,
along with radial velocity variations to check binarity \vs
pulsation.

For completeness, we note that we have also found one egregious large amplitude
variable in the database, \viz SC8~81586 (79.5626.10).  Despite the relative
sparsity of its OGLE and MACHO data and the noisiness of its MACHO data this
object has a sharply defined period of $P$ = 1.0589\thinspace d, and relatively
large amplitudes ($A_R$ = 0.024, $A_I$ = 0.025, $A_B$ = 0.036).  It is depicted
as a green diamond in the upper left side in Fig.~\ref{1}.

\begin{figure*}
\epsscale{0.9}
\epsscale{0.8}
\begin{center}
{\plotone{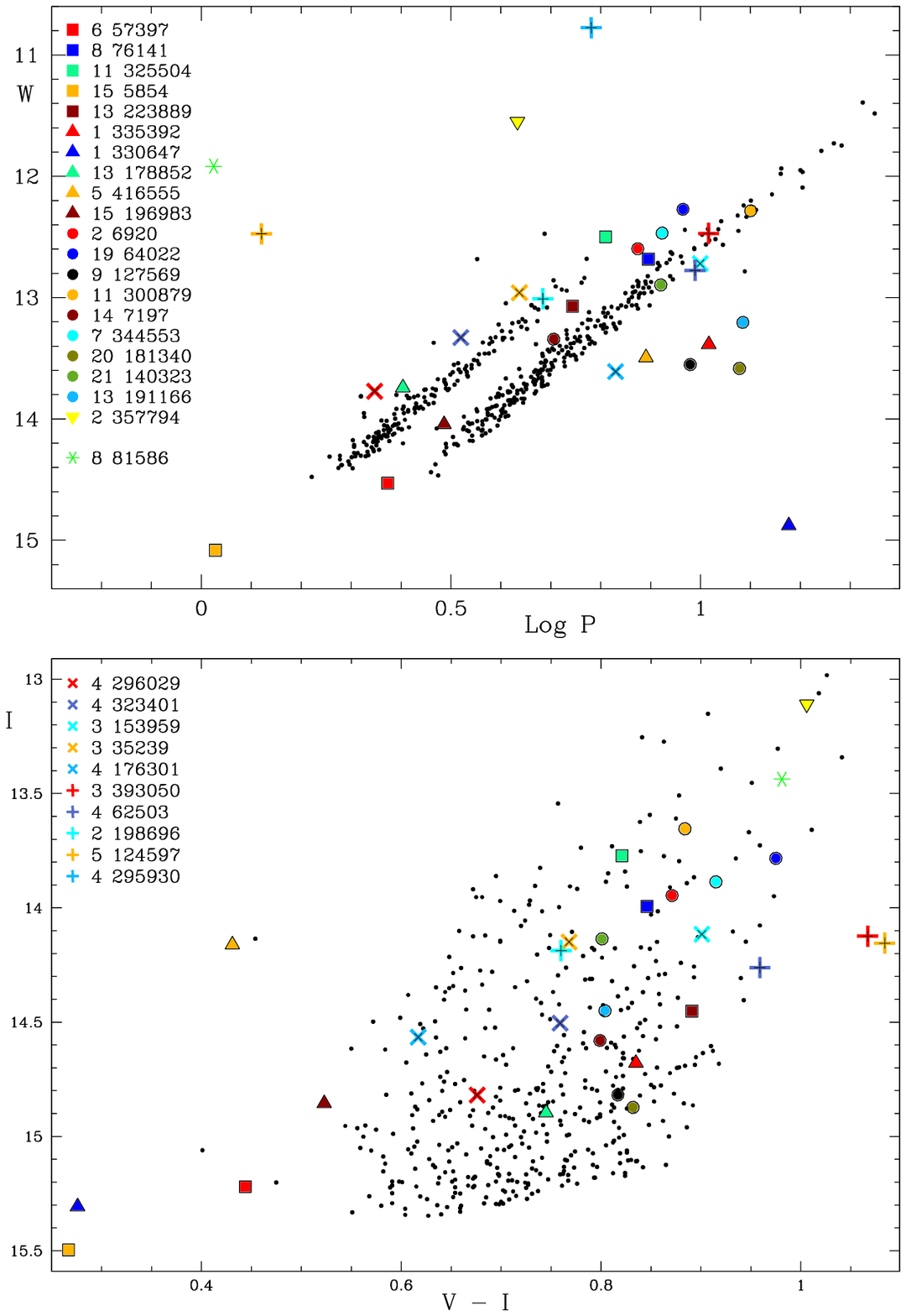}}
\caption{
Top: Period--$W$ plot for ULA variables (large symbols), where 
$W =I - 1.55\thi (V-I)$.  Also shown are the classical Cepheids (small black
dots) from our combined MACHO-OGLE data base.  Bottom: I, V--I plot for the
same stars.  The black dots represent the classical Cepheids.  The 5 squares,
 5 triangles and 9 filled cicles 
mark the position of the ULA Cepheids, the upside-down triangle
that of the potential strange Cepheid candidate SC2~357794.
The crosses and
Xs denote the objects from \cite{bwks05}.  SC4~295930 has a  V--I = 1.38,
outside the plotted range.  The green asterisk in the color-magnitude
diagram locates the egregious variable star SC8~81585 that is located in the
upper left corner in Fig.~\ref{1}.
}
\end{center}
\label{8}
\end{figure*}

\subsection{Fourier Spectra and Light Curves of the Singly Periodic Ultra-low
Amplitude (ULA) Objects}

\begin{figure*}
\begin{center}
\epsscale{0.9}
\epsscale{0.8}
{\plotone{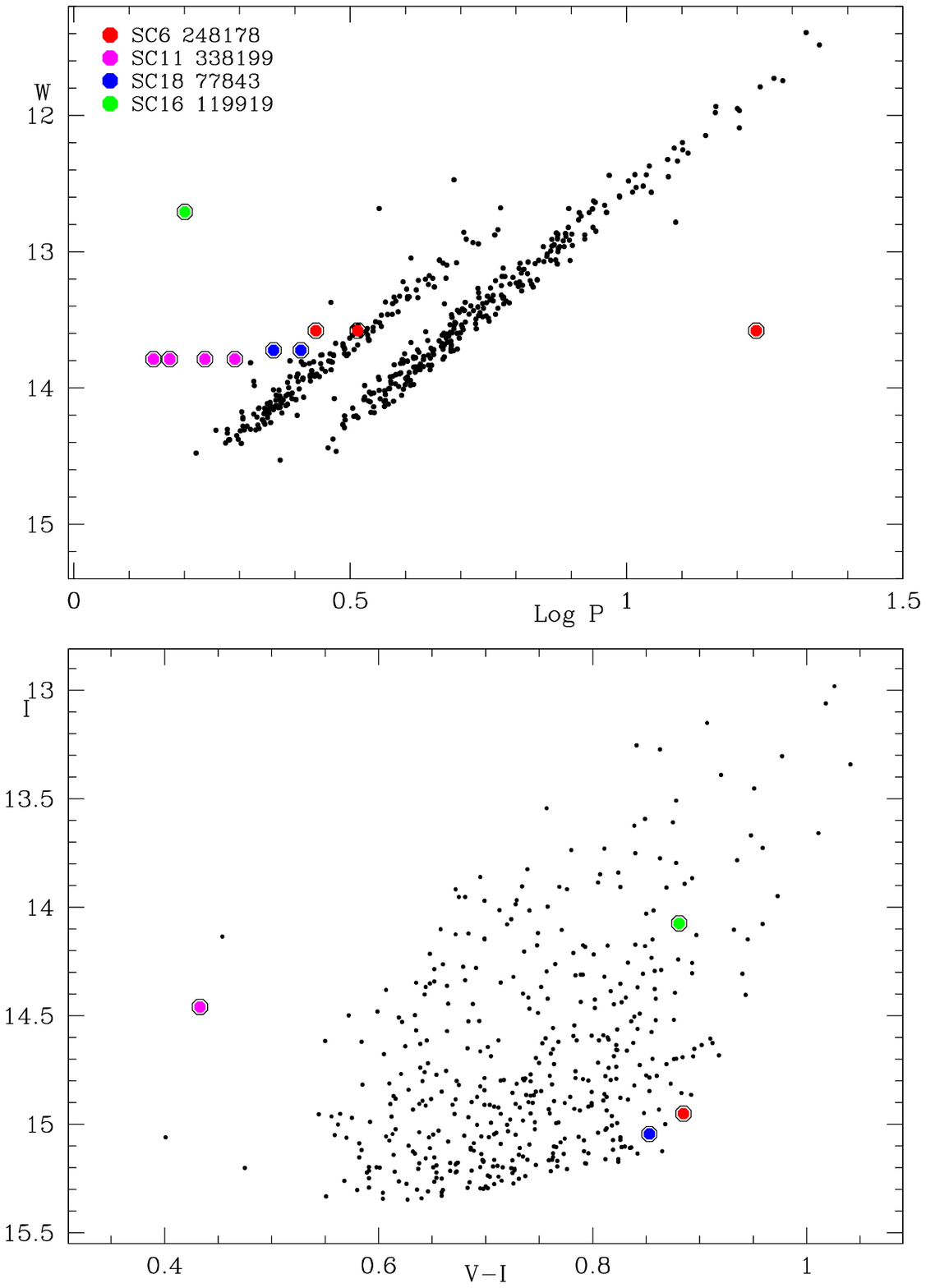}}
\caption{
Top:
Period--$W$ magnitude plot.  The objects of this paper are superposed on the
Cepheids from our combined MACHO-OGLE data base.  The various periods of each
star appear on a horizontal alignment; Bottom: (I,\,V--I)  CM plot.
}
\label{9}
\end{center}
\end{figure*}

\begin{figure*}
\begin{center}
\epsscale{0.67}
{\plotone{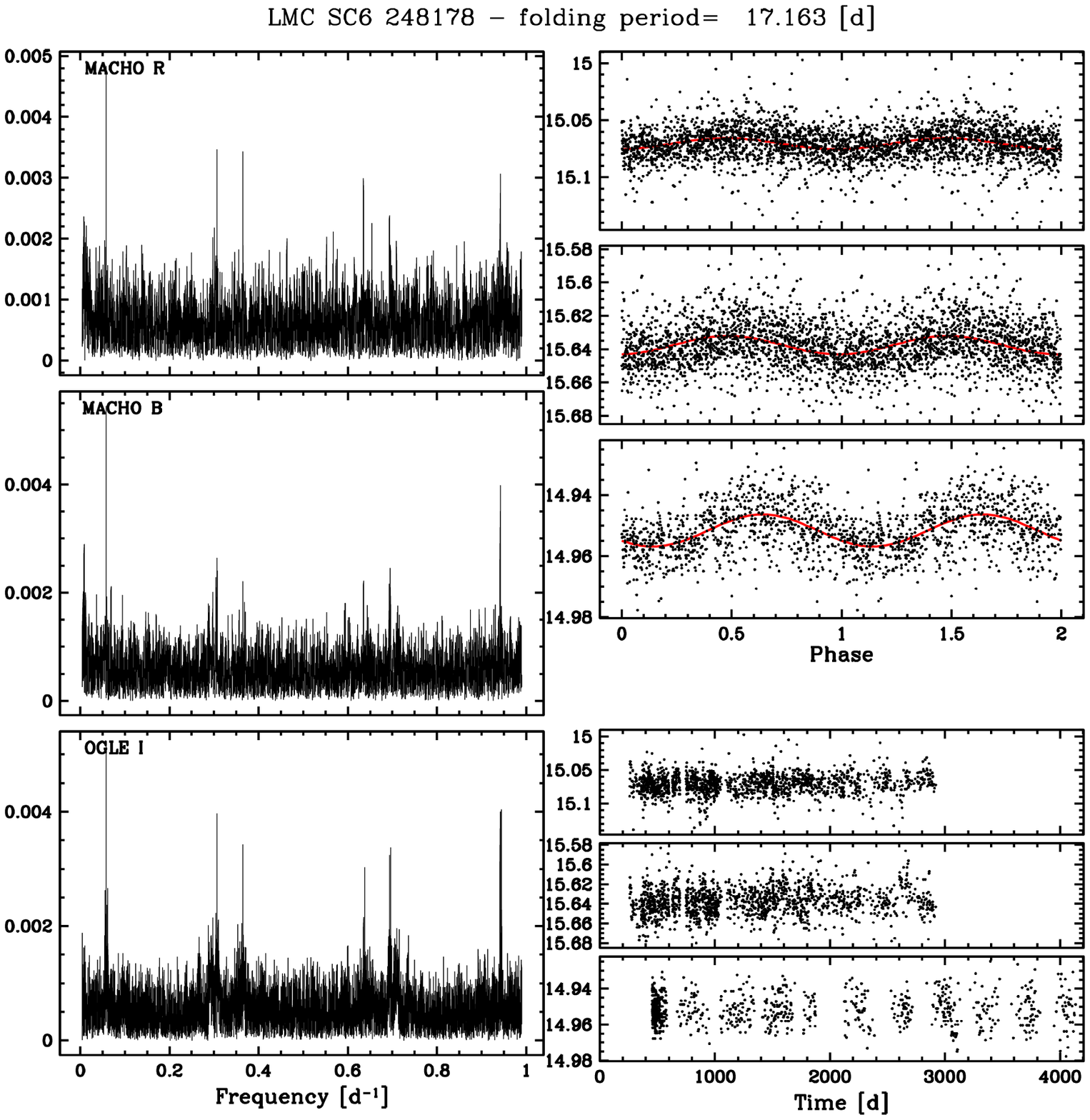}}
\caption{LMC SC6 248178 --
Left side, top down: 
Fourier amplitude spectra of the \MR, \MB\ and \OI\ data;
Right side, top: folded and phased data;
folding period appears on top;
bottom: corresponding data points.
}
\label{10}
\end{center}
\end{figure*}

\begin{figure*}
\begin{center}
\epsscale{0.67}
{\plotone{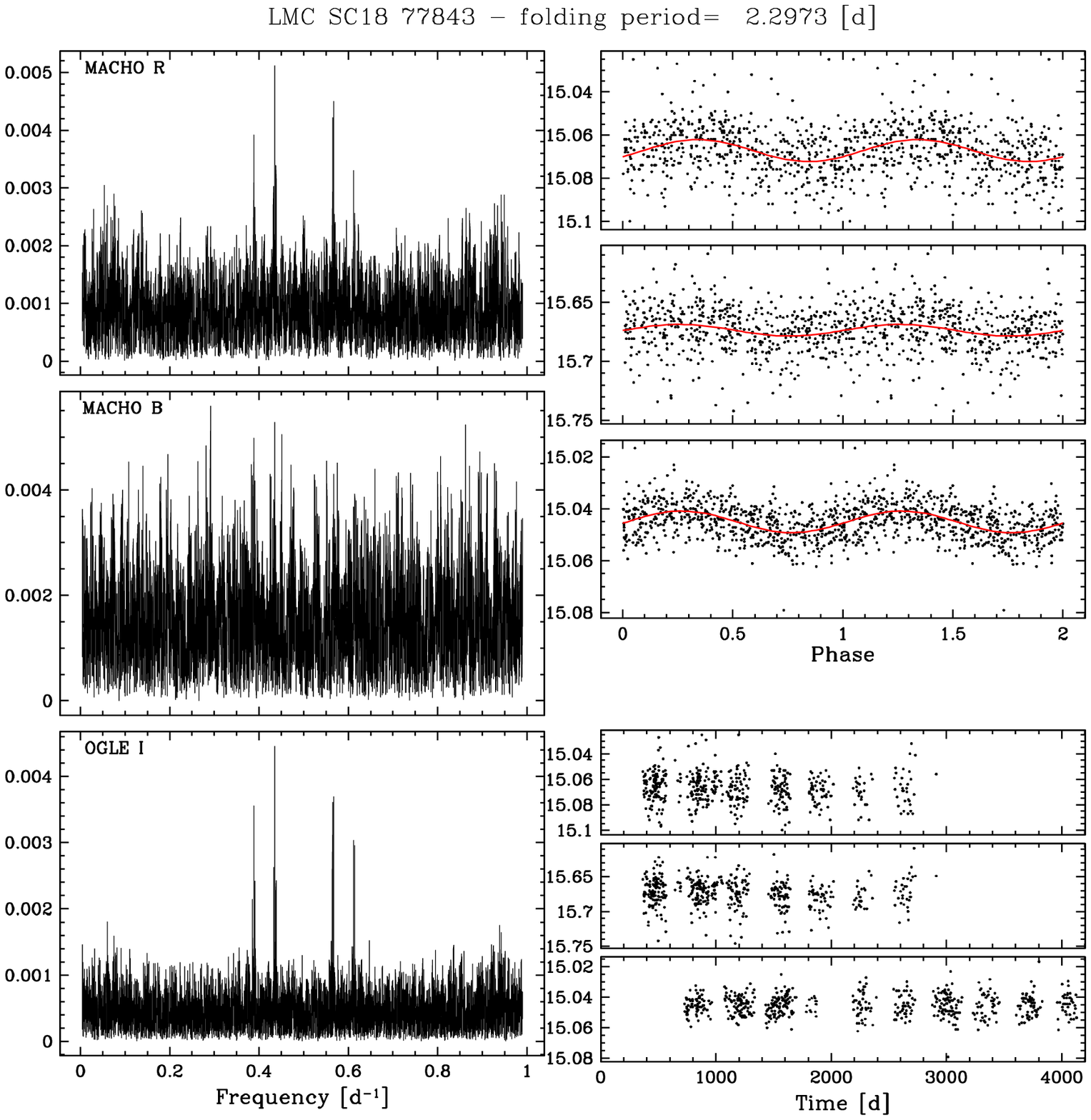}}
\caption{LMC SC18 77843 --
Left side, top down: 
Fourier amplitude spectra of the \MR, \MB\ and \OI\ data;
Right side, top: folded and phased data;
folding period appears on top;
bottom: corresponding data points.
}
\label{11}
\end{center}
\end{figure*}

\begin{figure*}
\begin{center}
\epsscale{0.67}
{\plotone{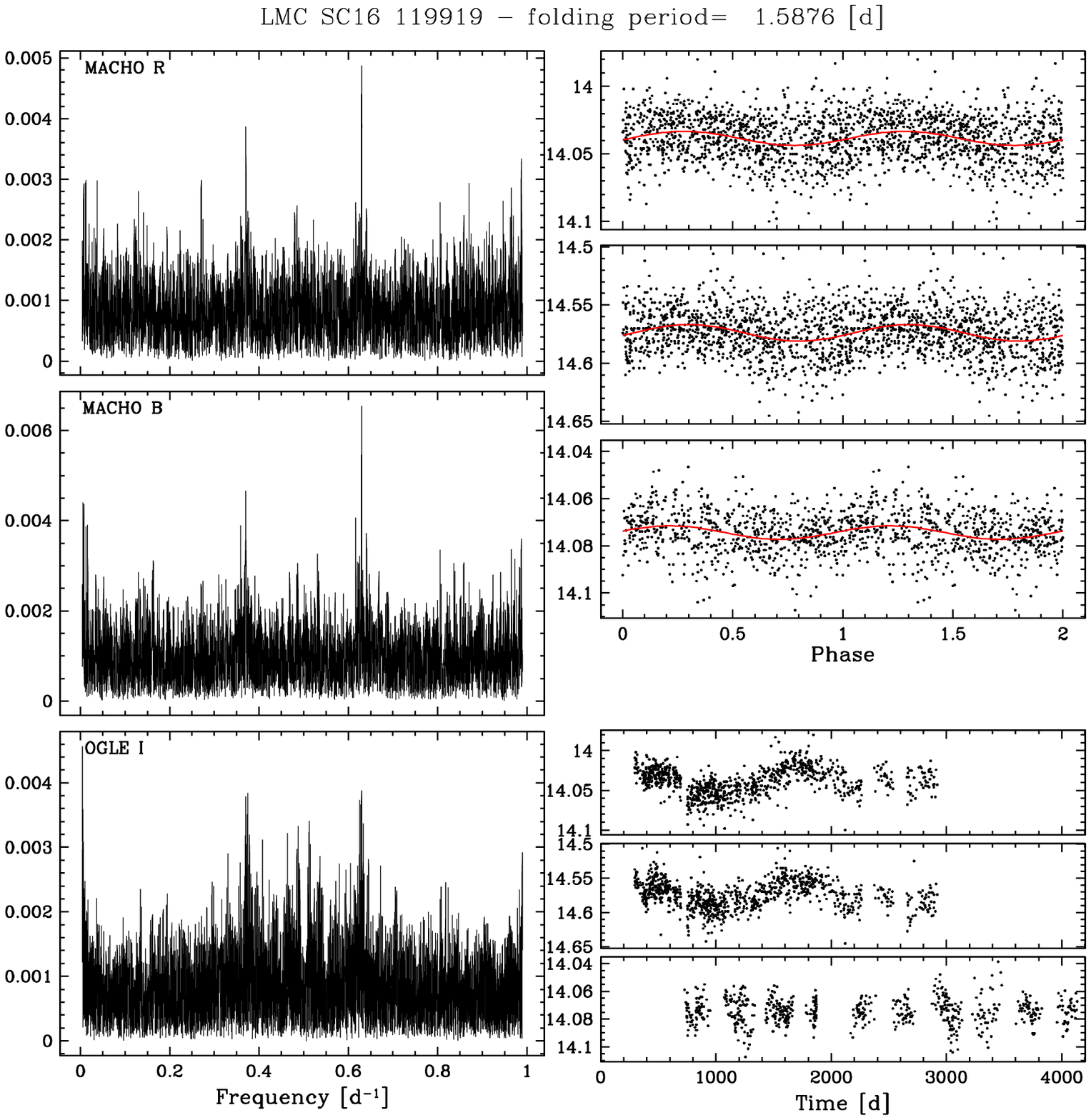}}
\caption{LMC SC16 119919 --
Left side, top down: 
Fourier amplitude spectra of the \MR, \MB\ and \OI\ data;
Right side, top: folded and phased data;
folding period appears on top;
bottom: corresponding data points.
}
\label{12}
\end{center}
\end{figure*}

\begin{figure*}
\begin{center}
\epsscale{0.67}
{\plotone{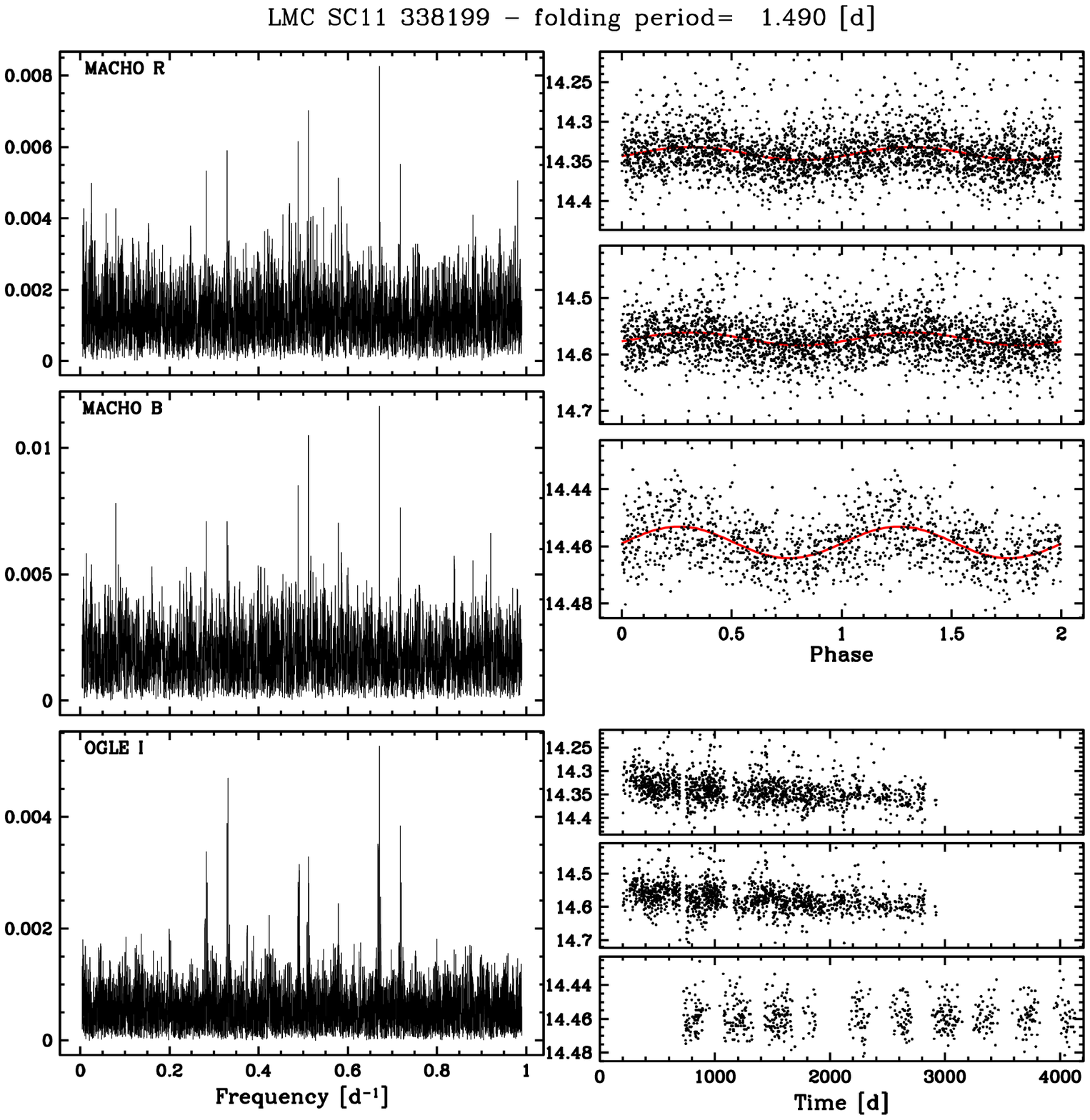}}
\caption{LMC SC11 338199 --
Left side, top down: 
Fourier amplitude spectra of the \MR, \MB\ and \OI\ data;
Right side, top: folded and phased data;
folding period appears on top;
bottom: corresponding data points.
}
\label{13}
\end{center}
\end{figure*}


Plots of the results of our Fourier analysis of each of these objects would
take up too much space in this paper.  We therefore give just a few
representative examples.

Our figures are in a standard format and contain:
(1)~Panels with the amplitude Fourier spectra 
for the \MR, \MB\ and \OI, respectively,
plotted for a frequency range of 0.004 to 0.99 \dinv.
(2) Panels with the actual data points on the bottom right;~
because the OGLE observations start three and 3/4 years (1376.5 days) after the
MACHO observations we have shifted them backwards in time in the OGLE panel for
better visualization.
(3) Panels on the top right side that display the light curves folded
with the period that is indicated on top of the figure;~  the MACHO and OGLE
data have been phased by taking the 1376.5\thi d shift into account.
The solid red curves in the top right panels represent single frequency
harmonic fits to the phased and folded \MR\, \MB, and \OI\ data.
In some cases we have made small adjustments to the folding period $f_R$ to
improve the phasing.

The analyses of SC8~76141, SC14~7197, and SC11~325504 are
exhibited in Figs.~\ref{3} -- \ref{5}.  They
are among the top 16 objects of Table~\ref{tab:ula} and they satisfy all our
criteria for variability, \viz clear peaks in all 3 bands, a strong 1 day
alias, good consistent phasing of the folded light curves.

Finally, we wish to give two examples of variability at the confidence limit.
Fig.~\ref{6} displays the analysis of SC13~178852.  The \OI\
data have a very convincing spectrum with a dominant peak at 0.3947\thi
d$^{-1}$.  While the corresponding peak is indeed the largest in \MR\ and \MB,
one would perhaps not have claimed variability on the basis of the MACHO or the
OGLE data alone.  However, it is unlikely that the {\sl same} object would show
common peaks, however weak if the variability were not real.  Note also that
the \MR\ spectrum shows the 1-f alias and that there is a strong hint of it in
the \OI\ spectrum.  Furthermore, the folded light curves show in-phase
variability.

In Fig.~\ref{7} we present perhaps our weakest case
(SC15~196983) for variability.  \OI\ and \MB\ have their dominant peak at
f$_{\rm o}$ =0.3261\thinspace d$^{-1}$.  \OI\ has a strong 1 day alias peak at
0.6739, which is also present in \MB\ albeit just barely.  However this is the
prominent peak in \MR\ where the f$_{\rm o}$ peak is hidden in the 'grass'.  We
have chosen the f$_{\rm o}$ peak as the actual frequency of this object for two
reasons: first because it is the dominant peak in the \OI\ data which have the
cleanest spectrum, and second, because the other peak gives a bad phasing
between the folded MACHO and OGLE data.

\subsection{Discussion}

Since many of the objects lie in crowded fields the question arises as to the
reliability of our assignments.  The variability analysis and the periods that
have been determined are immune to contamination by neigboring objects unless
those objects have variability themselves or our object is part of a binary,
for example.

On the other hand, contamination by a neighboring object will affect the
magnitude and the color of our objects.  We have looked in some detail at the 3
objects SC4~62503, SC5~124597 and SC4~295930, for example.  The first two are
isolated objects, but the third one is a messy clump, so it is possible (but
not necessary) that the photometry of this last object is poor.  This will
apply to both the OGLE and MACHO data.  V--I for this object is 1.38, way
outside the typical range (see the CM plot of Fig~8.  A further
look at SC4~295930 in I-\MR, V-\MB\ and V-I reveals that I is anomalously
bright, by at least 0.5 mag.  Consequently, it should be shifted down in the PM
plot and could well be a regular classical Cepheid of very low amplitude rather
than a strange Cepheid.

We now turn to Fig.~8.  The top panel is a blowup of the PM
($\Log P$, $W$) plot of Fig.~\ref{1}, and the bottom panel is
the corresponding CM (V--I, I) plot.  All the objects that are listed in
Table~\ref{tab:ula} and in \cite{bwks05} are identifiable in the figure.  The
squares, diamonds, filled circles and upright triangles indicate the newly
found ULAs.  The filled circles locate the potential foreground objects.  The
Xs denote the 5 variables that are in common with the previous work of
\cite{bwks05} which concentrated on MACHO Field 77.  The crosses are their
additional variables that lie marginally outside the cut that defines this
paper's data base (see \S1).  For reference we have again plotted the classical
Cepheids as small black dots.

These ULA objects are certainly a mixture of pulsating stars with perhaps a few
ellipsoidals that can also have amplitudes in the millimag range.  Of course,
the ULA Cepheids have to satisfy the known PM relations, but it is difficult to
distinguish between these two types of low amplitude variability.

One notes that 20 objects of Table~\ref{tab:ula} fall into the classical
Cepheid range, in both the PM and the CM diagrams.  Most of the ULA objects fall
outside of or near the edges of the F and O1 instability strips.  That ULA
Cepheids would be at the edges of the instability strip is consistent with
theory \citep{bk02,bwks05}.

However one might claim that the figure indicates that the ULA objects form two
separate, parallel fundamental and overtone PM relations in which case they
would be physically distinct from the regular Cepheids\thi (perhaps on the
first crossing of the instability strip rather than on the second or third
crossings on the blue loop).  The color-magnitude diagram, unfortunately has
too much scatter to provide any discriminating information.

One object, SC15~5854, with a period of 1.0674\thi d lies in the continuation
of the Cepheid PM relation and falls into the CM range of the Cepheids,
albeit quite beyond the tip.

There is a cluster of 6 stars that falls below the Cepheid PM relation with
periods 2--3 times longer than that of classical Cepheids of the same
magnitude.  Three of them are possibly foreground stars, as discussed in
\S2.  Again, we cannot be sure that these objects are not ellipsoidals,
but it seems unlikely because of the clustering in PM.  Furthermore, 4 of them
(SC9~127569, SC20~181340, SC1~335382, and SC13~191166) also cluster in CM.
This suggests that they are pulsators rather than ellipsoidals.  What is their
nature?  They are unlikely to be classical Cepheids because they have too small
a brightness for their period, and potential contamination by neighboring
objects would move them even lower in the PM diagram.  It is possible that
these stars are also ULA Population II stars.  However, SC5~416555, which is
very blue is perhaps an ellipsoidal.

The star SC2~357794, marked with an inverted triangle in Fig.~8,
falls into the proper CM range, but it has a period a factor of $\sim$5 smaller
than the fundamental Cepheid period, \ie the period the star would have if it
pulsated in the F mode.  The theoretical work of \cite{byk97} has shown that
some Cepheids can pulsate in a surface mode with a period that is about a
factor of 4--5 shorter than the corresponding fundamental period.  This object
therefore is possibly a strange Cepheid \citep{bk01}.  But one cannot rule out
that the light curve might be contaminated by a bright nearby star in which
case it could be a regular ULA Cepheid as we already suggested for object
SC5~124597 that \cite{bwks05} had identified as a strange Cepheid candidate.

Object SC1~330647 is a little offset with $W \sim$ 15 and $\Log P \sim
1.2$ and lies in the W~Vir instability region (see Fig.~\ref{1}.
It could be a ULA W~Vir star although its V--I color is very blue.

We conclude that most if not all of the ULA singly periodic objects in the data
set are Pop. I and Pop. II Cepheids, with the caveat that some might be
foreground stars or ellipsoidals.


\subsection{Other, Larger Amplitude Variable Objects}

We now turn to the larger amplitude singly periodic stars that are exhibited in
Table~\ref{tab:popII}.  The objects that we identify as probable binaries are
marked in orange and the probable Pop. II or RV~Tau in red in
Fig.~1.

Their identifiers, their average I magnitudes, Wesenheit magnitudes $W$ and
their periods (or cycling times) are given in Table~\ref{tab:popII}.  The
first two columns give the MACHO and OGLE names.  All these stars have I~band
Fourier amplitudes in the range 0.01 to 0.25.  An exception is the ULA
SC3~201554 with harmonic peak amplitudes of 0.0065 and 0.0075, that is shown in
yellow.  The objects that are displayed as triangles have regular light curves.
One object, SC4~53483, has a 'crested top' light curve that is typical of some
W Vir stars.  The remainder show behavior with cycle to cycle alternations.
These alternations are regular (squares) for some and some semi-regular
(circles).

The period association is sometimes ambiguous by a factor of 2 when
alternations are present.  We have not chosen the period with the highest
Fourier peak, but somewhat arbitrarily the one that gives visually the 
folded light curve with the least scatter.

One notes  that all but 2 points would fall on 2 parallel, relatively tight
PM relations that are separated by 0.3 (factor of 2 in P).  The binaries
(orange) fall predominantly on the right sequence, and the Pop. II/RV~tau stars
(red) predominantly on the left one.

Actually, from a physical point of view, for the pulsating stars the shorter
period is more relevant as it corresponds to the linearly excited pulsational
mode, whereas the longer period arises through period doubling that is caused
by nonlinear effects (\cite{bk86}, \cite{kb87}).

For the binaries, the reverse situation holds.  Here there are also two
sequences separated by 0.3 in $\Log P$ because the "no alt" stars can adopt a
period equal to half the orbital period (there are two maxima per orbit).

But, why does the binary sequence exist?  We will argue that it is because
stars on the blue loops have a certain size at a given luminosity and color,
and if they have a companion that is capable of causing ellipsoidal variations,
it must be at a small multiple of the stellar radius.  This radius increases
with luminosity (I) and color (V-I).  The conversion to $W$ will tend to take
out the V-I variation along tracks at given luminosity.  If we remember that
for both orbits and pulsation, $P \sim R^{1.5}$, and that the period of the
fundamental pulsation mode is not too different from the orbital period at the
stellar surface, we can see that qualitatively the binary sequence should be
parallel to the pulsation sequence: more detailed computations would be
needed to estimate the exact shift in $\Log P$.  The stars on the binary
sequence should therefore be in the LMC.

\begin{table*}
\caption{\small Larger Amplitude Singly Periodic Objects}
\noindent Columns:  OGLE and MACHO IDs, I band mag, period, 
I band Fourier amplitude, star type.
In the case of alternating cycles, 
the period has been chosen to visually 
give the best folded light curve, rather
than to correspond to the Fourier peak with the highest amplitude. 
\vspace{-0mm}
\begin{center}
\medskip
\begin{tabular}{l l l l l l l l}
\hline\hline
\noalign{\smallskip}
OGLE ID & MACHO ID  &    $M_I$  &    $W$  &        P[d]  & $A_I$ & &Type of star\\   
\noalign{\smallskip} 
\hline   
\hline
SC21 85305  &     6.6575.25  & 15.065 & 14.071  & 10.99 &       &  alt    & Pop. II\\
 SC8 52612  &    79.5622.20  & 14.638 & 13.835  & 10.99 & 0.038 &  no alt & Pop. II\\
 SC7 295173 &    78.6223.44  & 15.036 & 14.435  & 17.74 & 0.037 &  no alt & Pop. II\\ 
 SC4 53483  &    77.7306.73  & 14.902 & 14.133  & 20.21 &       &  crested & Pop. II\\
 SC4 287444 &    77.7426.59  & 15.288 & 14.911  & 26.97 &       &  alt    & binary\\
SC1  325044 &    81.8758.27  & 15.533 & 15.170  & 28.06 & 0.015 &  no alt & binary\\
 SC6 462158 &    80.6710.17  & 14.113 & 13.226  & 30.14 & 0.014 &  no alt & binary\\
SC10 137113 &   2.4902.4632  & 15.175 & 14.633  & 32.22 &       &  alt    &\\
SC10 105184 &    79.4896.51  & 15.225 & 14.478  & 35.61 &       &  alt    & Pop. II\\
 SC5 92493  &  78.6949.2259  & 14.090 & 12.989  & 39.4  &       &  semireg & Pop. II\\
 SC3 274381 &     77.7911.4  & 14.105 & 12.963  & 40.4  &       &  semireg & Pop. II\\
SC14 57979  &    19.3694.19  & 14.662 & 13.441  & 63.40 &       &  alt    & binary \\
SC11 338244 &  79.4659.3417  & 14.972 & 13.854  & 69.82 &       &  alt    & binary\\
SC11 338203 &    79.4780.27  & 14.569 & 13.306  & 71.7  &       &  semireg & Pop. II\\
 SC9 263456 &    79.5504.13  & 14.684 & 13.267  & 71.78 & 0.057 &  no alt & binary\\
 SC6 40874  &  78.6461.2152  & 14.361 & 13.385  & 72.76 &       &  alt & Pop. II\\
 SC8 181816 &  78.5856.2363  & 14.383 & 13.197  & 81.99 &       &  alt  & binary\\
SC14 100796 &     1.3810.19  & 14.038 & 13.382  & 88.95 &       &  alt & binary\\
 SC9 237168 &    79.5501.13  & 13.845 & 12.706  & 96.90 &       &  alt & binary\\
SC14 100798 &     1.3810.30  & 14.323 & 13.226  &101.26 &       &  alt & binary\\
\noalign{\smallskip}                   
 SC3 201554 &    77.7794.46  & 15.527 & 14.9457 & 25.57 & 0.007 & binary \\
\noalign{\smallskip}
\noalign{\smallskip}
\hline
\hline
\end{tabular}
\label{tab:popII}
\end{center}
\end{table*}

We have found low amplitude RV~Tauri and Pop. II
Cepheids in the LMC that do not appear in the catalogs of
\citep{macho99,ogle}.  We also find that, curiously, a set of binaries appear
to lie on essentially the same PM relation as the RV~Tauri/Pop. II Cepheids.

\section{Multiperiodic variables}

\begin{table*}
\caption{\small ULA Multimode stars}
\noindent Columns: OGLE and MACHO identifiers, I magnitude, V--I color,
significant frequencies, corresponding \MR\ Fourier amplitudes.
(The last object's amplitude is a little larger than our ULA criterion). 
A dagger in the last column denotes potential proper motion.
\vspace{-0mm}
\begin{center}
\medskip
\begin{tabular}{l l c l c c c c c c c c c c c}
\hline\hline
    \noalign{\smallskip}
OGLE & MACHO & I & (V-I) & $f_1$ & $f_2$ & $f_3$ & $f_4$ & $A_1$ & $A_2$ &  $A_3$ & $A_4$ \\
SC6 248178 & 78.6699.27  & 14.951 & 0.885 & 0.05819 & 0.36484  & 0.30657 & & 0.0048 & 0.0034 & 0.0034 & \\ 
 SC18 77843 & 11.9838.21 &15.045 & 0.853 & 0.43528 & 0.38817 & & &  0.0050 & 0.0039 & & \\ 
SC16 119919 & 81.8998.12 & 14.074 & 0.881 & 0.63018  & 0.00072  & & & 0.0048 & 0.0159 & &  \\
SC11 338199 & 79.4659.3384 & 14.459 & 0.433 &  0.67103 & 0.51146 & 0.71783  & 0.57911 & 0.0083 
               & 0.0070 & 0.0055 & 0.0051 & $\dagger$\\
\noalign{\smallskip}
\hline
\hline
\end{tabular}
\label{tab:mper}
\end{center}
\end{table*}

Four of the ULA stars have multiple periods.  They are displayed in the PM plot
of Fig.~\ref{9}.  Again for reference we have superposed the Cepheids
as small black dots.  The bottom panel displays the corresponding (I,\,V-I)
plot.  The properties of these objects appear in Table~\ref{tab:mper}.

The results of the individual Fourier analyses are presented in
Figs.~(\ref{10}, \ref{11},
\ref{12} and \ref{13}). 



\vskip 5pt 
* LMC SC6 248178:\quad The results of the analysis are displayed in
Fig.~\ref{10}.  Table~\ref{tab:mper} indicates that this star
has its largest frequency peak at $f_1$ = 0.05819\dinv, \ie 17.185\thi d, and
two close frequencies, $f_2$ = 0.36484 and $f_3$ = 0.30657\dinv.  One notes
that $f_3\sim f_2+f_1$.  Even though the 3 amplitudes are all in the low
millimag range, the 3 peaks stand out very clearly in \MR, \MB\ and \OI.  Peaks
2 and 3 suggest that if this star is a Cepheid it may be a higher order
double-mode pulsator with a large period ratio of $P_3/P_2$ = 0.84028.  If the
long period of 17.185\thi d were due to an unseen binary companion that would
not explain the observed frequency lock.


\vskip 5pt
* LMC SC18 77843: \quad 
The analysis of this object appears in Fig.~\ref{11}.  
There are two peaks with a
period ratio of 0.89177 which is larger than that of a double
mode Cepheid that pulsates in the O1 and O2 modes.  The star also lies somewhat
to the left of the PM
relation, both when plotted with I and
with the Wesenheit magnitude $W$, consistent with the expected
positions of higher order modes.


\vskip 5pt
* LMC SC16 119919:\quad
This star is a 1.59\thi d period variable with an extremely long period
(1380\thi d) modulation (Fig.~\ref{12}).
Looking at the light curve points one would think that we have mismatched
the OGLE object with the wrong MACHO object. However the spectral peaks and the
phased, folded light curves suggest otherwise.

The 1.59\thi d period is much shorter than expected for a Cepheid
of similar magnitude and color.  Similarly, the long 1380\thi d
is hard to explain.  It could be a binary with a very bright companion,
although the amplitude of the long period variation 
would be very large for ellipsoidal motion.  The period ratio is also at odds
with that of the blue multi periodic variables that \cite{menni} discovered.
 

\vskip 5pt
* LMC SC11 338199: \quad
This star is multiperiodic with 4 frequencies that show up simultaneously in
\MR, \MB\ and \OI\ (Fig.~\ref{13}).  No other peaks seem to be
significant.  All four periods are consistent with low order radial pulsation
modes, but it is not obvious why they they should all be excited for a star of
this brightness.  However, there is always the possibility that the light
curves of this object are contaminated by a neighboring star, in which case it
would fall into a brightness regime where one might expect AI~Velorum type
multimode pulsation.

There is recent evidence for proper motion in this star \citep[][unpublished OGLE-III results]{ogle08}.
If SC11~338199 is a foreground star that would make it even brighter and its
nature harder to explain.
 

\section{Summary}

Fourier analysis of the light curves of LMC stars in the very broad vicinity of
the Cepheid instability strip part yielded 25 new singly periodic and 4 new
multi-periodic variables with ultra-low amplitudes (ULA)\thi  ($\approxlt$ 0.01
mag).
The advantage of our approach has been that, for many of these objects, it would
be difficult to claim variability on the basis of either the MACHO or 
the OGLE data only.
In addition, we have found RV~Tauri and Pop. II Cepheids that do not
appear in either the MACHO or OGLE catalogs.  

If we include the objects of \cite{bwks05} we have 20 ULA objects that fall
close to the fundamental (F) and overtone (O) Cepheid PM relations.  From
Fig.~8
it appears that these stars might form separate
sequences that are slightly above and parallel to the classical F and O LMC
Cepheids.  If this effect turns out to be real it will pose an interesting
challenge to explain the nature of these objects.  There is evidence
that 9 of these ULA stars show proper motions and that they could
be foreground stars, but that raises other questions as to their nature.  Even
if we accept them as foreground this
still leaves enough (14 instead of 20) objects to suggest separate PM
relations.

Combining this and previous work \citep{bwks05} we now have 3 stars that are
strange Cepheid candidates \citep{bk01}, although the light curve of one of them,
SC4~295930, could be contaminated by a very bright red star which would place
it erroneously above the PM relation.

Six closely clustered objects fall below the classical Cepheid PM relation (see
Fig.~8).  They could be ULA Pop. II Cepheids or
ellipsoidals, although 3 of them could be foreground stars.  
A further, very blue object, SC1~330647, could be a ULA W~Vir star.
Interestingly, a set of
binaries appear to lie on essentially a PM relation parallel to that of the
RV~Tauri/Pop.~II Cepheids.
Finally, we have uncovered 4 objects that show multiple modes of ultra-low
amplitude.

\acknowledgments

It is a great pleasure to thank Zolt\'an Koll\'ath for providing us with his
MUFRAN software.  This work has been supported by NSF (AST07-07972 and
OISE04-17772) at UF.  JRB gratefully acknowledges the hospitality of Mount
Stromly Observatory where this work was started.
We wish to thank an anonymous referee for his comments which led to a much
improved paper.


\end{document}